\documentclass[lettersize,journal]{IEEEtran}

\usepackage{lipsum}
\usepackage{array}
\usepackage[caption=false,font=normalsize,labelfont=sf,textfont=sf]{subfig}
\usepackage{textcomp}
\usepackage{stfloats}
\usepackage{url}
\usepackage{verbatim}
\usepackage{cite}
\usepackage{amsmath,amssymb,amsfonts}
\usepackage{graphicx}
\usepackage{textcomp}
\usepackage{xcolor}
\usepackage[abs]{overpic}
\usepackage{algorithm}
\usepackage[end]{algpseudocode}
\usepackage{etoolbox}
\usepackage{subcaption}
\usepackage{hyperref}
\usepackage[inline]{enumitem}
\usepackage{flushend}
\usepackage{pgfplots}
\usepackage{filecontents}

\def\BibTeX{{\rm B\kern-.05em{\sc i\kern-.025em b}\kern-.08em
    T\kern-.1667em\lower.7ex\hbox{E}\kern-.125emX}}
\usepackage{balance}
\begin{document}
\title{5G-Enabled Teleoperated Driving: An Experimental Evaluation}

\author{Mehdi~Testouri,
        Gamal~Elghazaly,
        Faisal~Hawlader,
        and~Raphael~Frank
\thanks{All authors are within 
SnT - Interdisciplinary Centre for Security, Reliability and Trust, University of Luxembourg. 
29 Avenue John F. Kennedy, 1855 Luxembourg, 
e-mail: \{mehdi.testouri, gamal.elghazaly, faisal.hawlader, raphael.frank\}@uni.lu}}

\maketitle

\begin{abstract}
Teleoperated driving enables remote human intervention in autonomous vehicles, addressing challenges in complex driving environments. However, its effectiveness depends on ultra-low latency, high-reliability communication. This paper evaluates teleoperated driving over 5G networks, analyzing key performance metrics such as glass-to-glass (G2G) latency, RTT and steering command delay. Using a real-world testbed with a Kia Soul EV and a remote teleoperation platform, we assess the feasibility and limitations of 5G-enabled teleoperated driving. Our system achieved an average G2G latency of 202ms and an RTT of 47ms highlighting the G2G latency as the critical bottleneck. The steering control proved to be mostly accurate and responsive. Finally, this paper provides recommendations and outlines future work to improve future teleoperated driving deployments for safer and more reliable autonomous mobility.
\end{abstract}


\section{Introduction}
\label{sec:Introduction}

The advent of autonomous vehicles has revolutionized the transportation industry, promising safer, more efficient and sustainable mobility solutions \cite{zhao2024remote}.
However, fully autonomous systems continue to encounter significant challenges in handling complex and unpredictable driving environments, including construction zones, adverse weather conditions, or ambiguous traffic situations \cite{georg2018teleoperated}. 
Teleoperated driving offers a viable solution to these limitations by enabling remote human control in challenging situations \cite{tang2014teleoperated}. 
This approach combines the benefits of human decision-making with the precision and responsiveness of automated systems, ensuring safer and more reliable operations \cite{10178441}.

The effectiveness of teleoperated driving depends mainly on G2G latency \cite{g2g}, the total delay from event capture by the vehicle sensors to its display at the operator control station. 
This latency is influenced by sensor acquisition, data encoding and decoding, network transmission, and display rendering. Maintaining a low and stable G2G latency is critical for real-time situational awareness, precise vehicle control, and driving safety.
The successful implementation of teleoperated driving relies heavily on the underlying communication infrastructure, which must provide ultra-low latency, high reliability, and sufficient bandwidth to transmit real-time sensor data, video feeds \cite{hawlader2024leveraging, hawlader2025cloud}, and control commands \cite{testouri2025robocar}. 
Conventional 4G/LTE networks suffer from high jitter, packet loss, and network congestion, making them unable to meet the requirements for teleoperated driving.
However, the rapid deployment of 5G has gained significant attention for its potential to enhance teleoperated driving by reducing latency, improving reliability, and ensuring seamless connectivity. 
Despite the promise of 5G for teleoperated driving, real-world evaluations remain limited, with key challenges unresolved.  
Most research relies on simulations \cite{cislaghi2023simulation}, creating uncertainties about network performance in real-world teleoperated driving \cite{saez2023design}. 
Bridging this gap is essential to assess whether 5G can provide the reliability and responsiveness required for large-scale teleoperated driving deployment.
This paper presents an experimental evaluation of teleoperated driving over 5G networks, focusing on the interplay between network performance and driving safety. 
We use standard metrics, including G2G latency, round-trip time (RTT) for driving command transmission, network jitter, steering dynamic response induced delay, and the impact of overall latency on steering angle stability for safe teleoperated driving operation. 

\begin{figure}[t!]
    \centering
    \includegraphics[scale=0.44, trim= 0cm 0cm 0cm 0cm,clip]{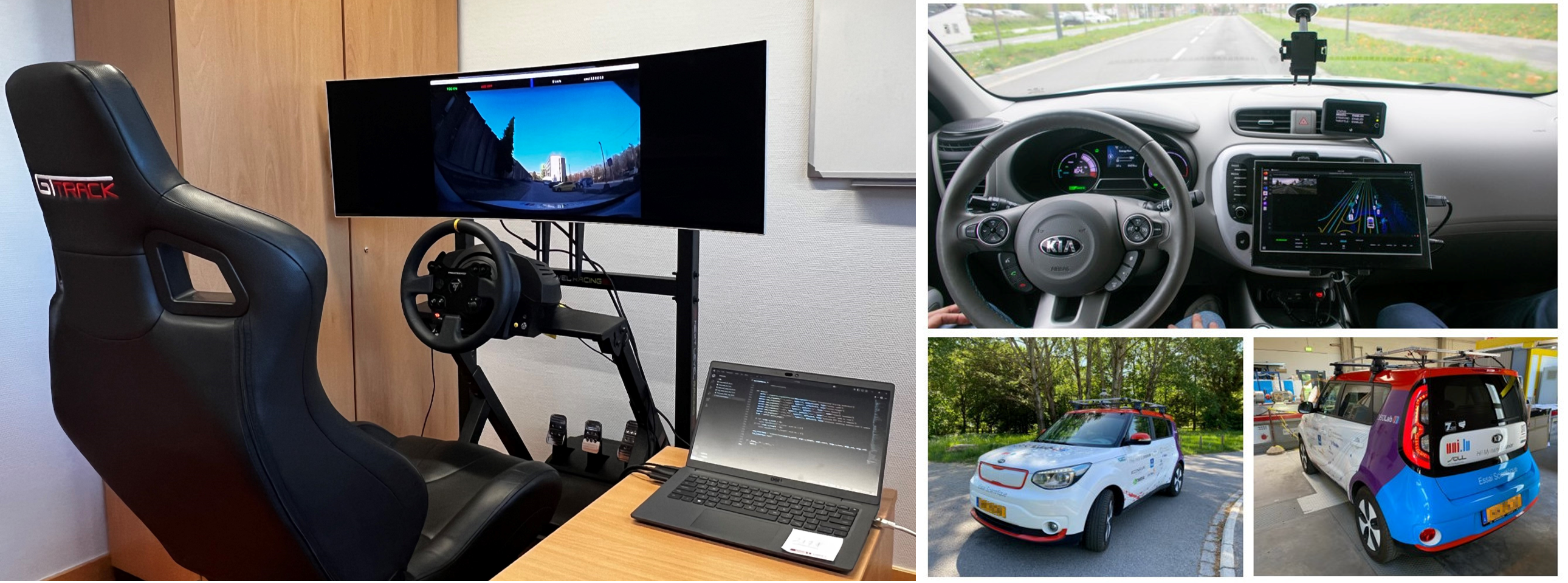}
    \caption{Experimental vehicle based on Kia Soul EV with the teleoperated driving interface.}
    \label{fig:enter-label}
\end{figure}
Our key contributions are summarized as follows:
\begin{itemize}
    \item \textbf{Design and Implementation of Teleoperated Driving Framework:} We have designed and implemented a comprehensive teleoperated driving framework. The system utilizes an autonomous vehicle equipped running Robocar, our open-source autonomous driving software (ADS) \cite{testouri2025robocar}. It also integrates a 5G communication network and a remote teleoperation interface, enabling precise vehicle control and experimentation in real-world operational conditions.
    \item \textbf{Experimental Evaluation and Performance Analysis:} Our study rigorously evaluates the capability of the system to support teleoperated driving by measuring standard evaluation metrics such as G2G latency, RTT, packet loss and jitter. We also analyze the overall steering delay induced by the dynamic response of the steering system and the network latency.
\end{itemize}

The remainder of this paper is organized as follows. Section~\ref{sec:related-works} reviews related work on teleoperated driving. Section \ref{sec:system-design-and-specifications} describes our teleoperated driving setup and its specifications. Section \ref{sec:software-architecture} presents the software architecture of our teleoperated driving platform. Section \ref{sec:experiments} presents the experimental results and analysis. Finally, Section \ref{sec:conclusion} concludes the paper and outlines directions for future research.

\section{Related Works}
\label{sec:related-works}

\begin{figure*}[t!]
    \centering
    \includegraphics[width=0.99\linewidth]{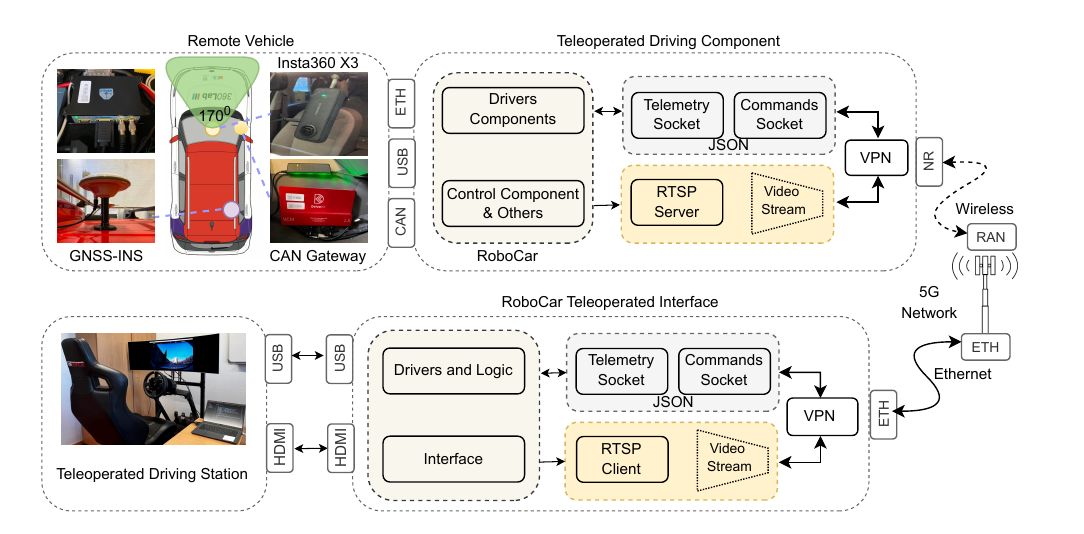}
    \caption{Description of the hardware and software architecture of the experimental setup for teleoperated driving.} 
    \label{fig:tod_arch}
\end{figure*}

Teleoperated driving has emerged as a crucial solution to bridge the gap between fully autonomous and human-driven vehicles \cite{tang2014teleoperated}, particularly in complex or unpredictable environments where autonomous vehicles struggle to make safe decisions \cite{kloppel2023identifying}. 
Several studies have explored the design, implementation and evaluation of teleoperated systems \cite{tang2014teleoperated}, focusing on hardware \cite{neumeier2019towards}, software \cite{kim2025development, schimpe2022open}, simulation-based evaluation \cite{cislaghi2023simulation, zhao2024remote} and real dataset evaluation \cite{neumeier2019measuring}. 
Reliable teleoperation depends on several critical factors, as consistently demonstrated across various studies \cite{shahbazi2018systematic, passenberg2010survey, arcara2002control}. Foremost, network stability is paramount for effective operation, as highlighted by \cite{g2g}.
Additionally, ultra-low latency is essential for maintaining real-time control \cite{eusebi1998force}, a point underscored by \cite{schimpe2022open}. 
Equally important is seamless data transmission \cite{de2025bridging}, which significantly influences performance. 
Collectively, these elements present formidable challenges in achieving consistent real-time control, as discussed in \cite{7976319, prasad20235g}.
Existing research has shown that network congestion and latency spikes often disrupt teleoperation \cite{mahmood2023remote}, causing steering instability, delayed braking \cite{aliaga2004experimental}, and operator disorientation \cite{neumeier2019measuring, cislaghi2023simulation}. 
Even in controlled testbed environments \cite{neumeier2018way}, network fluctuations \cite{kloppel2023identifying}, jitter \cite{gnatzig2013system}, and packet loss continue to impact teleoperation performance \cite{schimpe2022open, cislaghi2023simulation}. 
While 5G networks promise to reduce these delays \cite{prasad20235g}, real-world evaluations remain limited. 
Furthermore, the effect of network impairments on operator control accuracy remains unexplored \cite{mahmood2023remote ,zhao2024remote}, making it difficult to define safe performance thresholds \cite{neumeier2019measuring}.
Despite progress in system development and network integration \cite{georg2018teleoperated}, most research still relies on simulations or controlled environments \cite{neumeier2019yet, kim2024teleoperated}, limiting the applicability of findings to real-world deployments. 
Additionally, there is a lack of comprehensive evaluations using real vehicles \cite{testouri2025robocar}, real networks \cite{saez2023design}, and realistic driving conditions \cite{kashwani2024evaluation}, highlighting the need for further empirical validation before large-scale adoption. 
The authors in \cite{aliaga2004experimental} demonstrated an experimental quantitative comparison of different control architectures for teleoperation. However they failed to analyze G2G latency and steering control.

This paper addresses these gaps by presenting an experimental evaluation of teleoperation over 5G networks. We analyze network latency, jitter and packet loss, in addition to G2G latency, commands RTT and control steering delay to assess their impact on operator control. Unlike previous studies that focus on simulated environments or theoretical models, we directly correlate network conditions with teleoperation accuracy, providing empirical insights for real-world deployments.
\section{System Description and Specifications}
\label{sec:system-design-and-specifications}
This section describes the hardware and system specifications of the setup used for teleoperated driving experiments. This includes the description and specification of the test vehicle, all sensors and equipment, as well as the teleoperation interface. 

\subsection{Remote Vehicle}
The test vehicle is a modified KIA Soul EV equipped for autonomous and teleoperated driving \cite{testouri2025robocar,varisteas2019junior}. The vehicle comes with an integrated drive-by-wire system for brake, acceleration and steering. To facilitate accessibility of CAN bus and provide a reliable takeover system, the vehicle is also equipped with a DriveKit from Polysync that is  based on the Open Source Car Control (OSCC) project \cite{oscc}. The DriveKit provides a customized CAN interface to control the vehicle brake, acceleration, and steering while maintaining a level of safety to allow the driver to takeover in case of emergency or system failure. To host autonomous and teleoperated driving functions, the vehicle is equipped with an onboard computer Sintrones ABOX-5200G4 with an Intel Core i7-8700T CPU, an NVIDIA GTX 1060 GPU and 32GB of RAM. Localization in the vehicle is based on a Trimble BX992, an all-in-one high-precision GNSS-INS capable of providing a centimeter-level positioning and a precise heading at 20Hz. The camera used for teleoperation is an Insta360 X3 offering a 170-degree field of view in single-lens mode. Connectivity in the vehicle is provided via a consumer-grade 5G-enabled smartphone. The network characteristics of the setup will be discussed in Section \ref{sec:experiments}. Fig. \ref{fig:tod_arch} shows the remote vehicle hardware setup described. 

\subsection{Teleoperation Interface}

The teleoperated driving interface is a state-of-the-art setup designed to deliver an immersive and realistic driving experience. At its core is a 49-inch ultra-wide curved monitor and dual QHD 5120$\times$1440 resolution, providing a panoramic field of view that mimics real-world driving conditions. This display is paired with a Next Level Racing GT Track Simulator Cockpit supporting a racing steering wheel from Thrustmaster with force feedback offering direct drive technology for ultra-realistic feedback. The simulator supports a set of three high-precision pedals featuring a load cell brake pedal and customizable resistance, also from Thrustmaster. To further enhance immersion, the simulator seat is mounted on a Motion Platform V3 from Next Level Racing, which replicates the physical sensations of driving, including acceleration, braking, and cornering forces. This teleoperated driving interface is ideal for professional training, recreational use, or anyone seeking a truly immersive driving experience. Fig. \ref{fig:tod_arch} shows the whole setup of the driving cockpit and all supported components.

\section{Software Architecture}
\label{sec:software-architecture}

This section provides a detailed overview of the architecture of the software components deployed both on the vehicle and within the teleoperation interface. It outlines the key modules, their interactions, and the overall system design.

\subsection{Remote Vehicle}

The remote test vehicle is equipped with RoboCar as the Autonomous Driving System (ADS) software. RoboCar is an in-house, modular, ROS2-based platform developed specifically for autonomous driving research, as outlined in \cite{testouri2025robocar}. For this work, a dedicated teleoperation component has been developed, designed to operate at 100Hz to minimize end-to-end latency during teleoperation tasks. As depicted in Fig.~\ref{fig:tod_arch}, this component utilizes two ZeroMQ \cite{Hintjens2013ZeroMQMF} TCP sockets. The first is for transmitting teleoperation commands, while the second is used for receiving vehicle telemetry data. Both the teleoperation commands and telemetry messages are serialized using the JSON format for efficient data exchange. 
For connectivity, a VPN server is configured to connect the remote vehicle to the teleoperated driving interface, both as VPN clients. Connectivity to the remote vehicle is provided via 5G cellular communication, providing reliable and high-speed connectivity necessary for real-time teleoperation. The teleoperated driving interface, on the other hand, uses an  Ethernet connection to the internet. For visual feedback, camera frames are captured and preprocessed in real-time before being transmitted to a local RTSP server \cite{schulzrinne1998real}. The video stream is encoded using H.264 compression through FFmpeg, ensuring high-quality video transmission at a consistent 30 frames per second. This setup facilitates smooth and responsive communication between the teleoperation interface and the vehicle, enabling seamless and reliable remote control operations.

\subsection{Teleoperated Driving Interface}
The teleoperated driving interface is implemented in Python and includes a compact dashboard that provides real-time information about the system status, along with critical telemetry data such as the remote vehicle speed and steering position. These details are received from the vehicle telemetry messages and are displayed for the operator's reference. The interface continuously reads input from the steering and pedals, and sends the corresponding normalized control commands to the remote vehicle in the form of a JSON message. To enhance user experience and system ergonomics, the steering and pedal inputs are carefully calibrated, ensuring that the interface is intuitive and responsive. In addition to the control interface, the vehicle camera stream is displayed in a separate window using an FFmpeg RTSP client. This allows the operator to have a clear, real-time view of the vehicle surroundings, ensuring better situational awareness during remote operation.

\section{Experiments} 
\label{sec:experiments}
This section covers the experiments conducted to assess the capabilities of the presented teleoperated driving system. We first assess the general network performance before measuring G2G latency, RTT and finally evaluating the dynamic performance of steering control system. 

\begin{figure}[t]
    \centering
    \includegraphics[scale=0.3, trim= 0cm 0cm 0cm 2cm,clip]{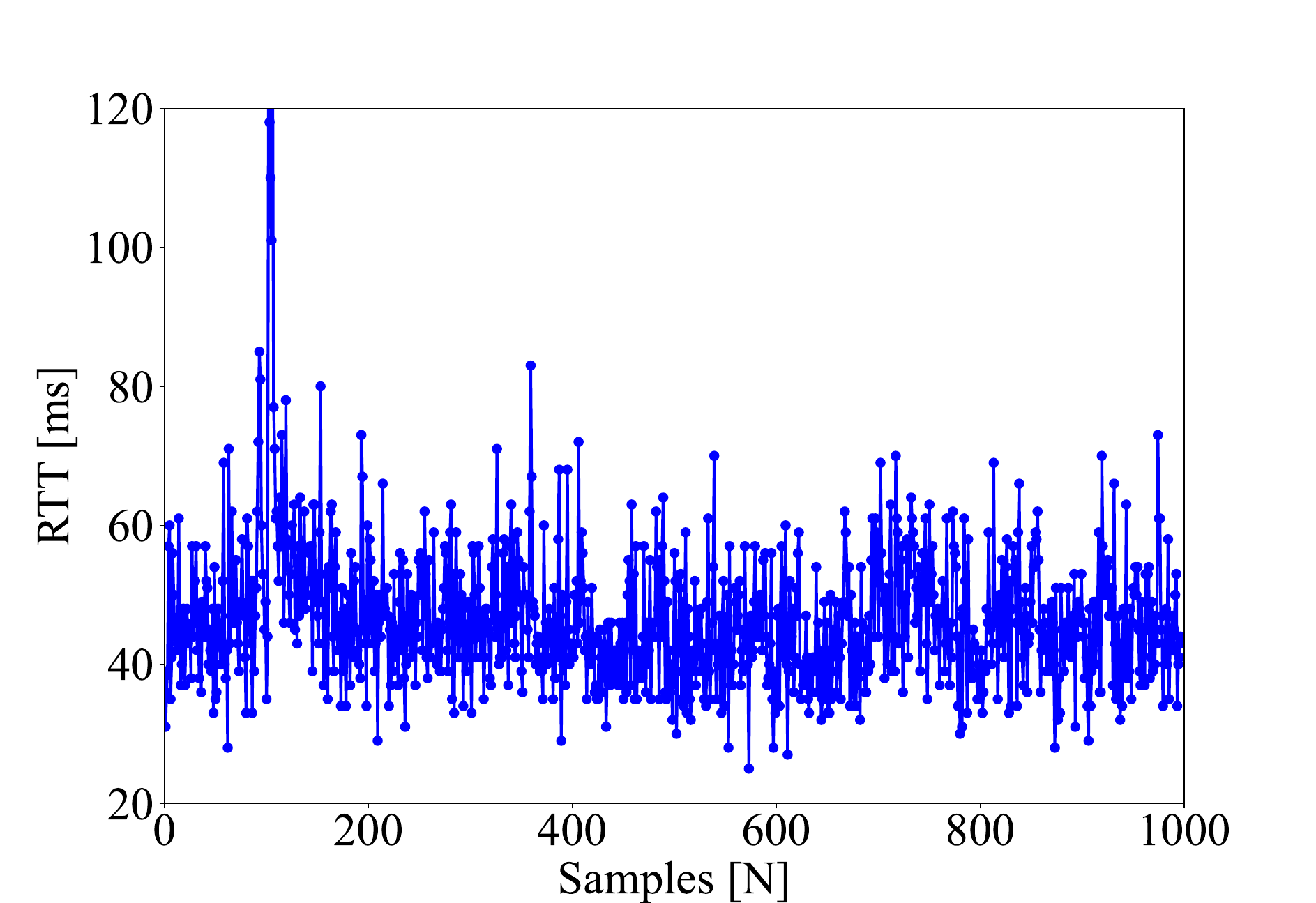}
    \caption{RTT measured as the time elapsed between a sent teleoperation command and a corresponding reply from the remote vehicle. The achieved RTT enables safe teleoperated driving at moderate to high speeds.}
    \label{fig:rtt}
\end{figure}

\begin{figure}[t]
    \centering
    \includegraphics[scale=0.3, trim= 0cm 0cm 0cm 2cm,clip]{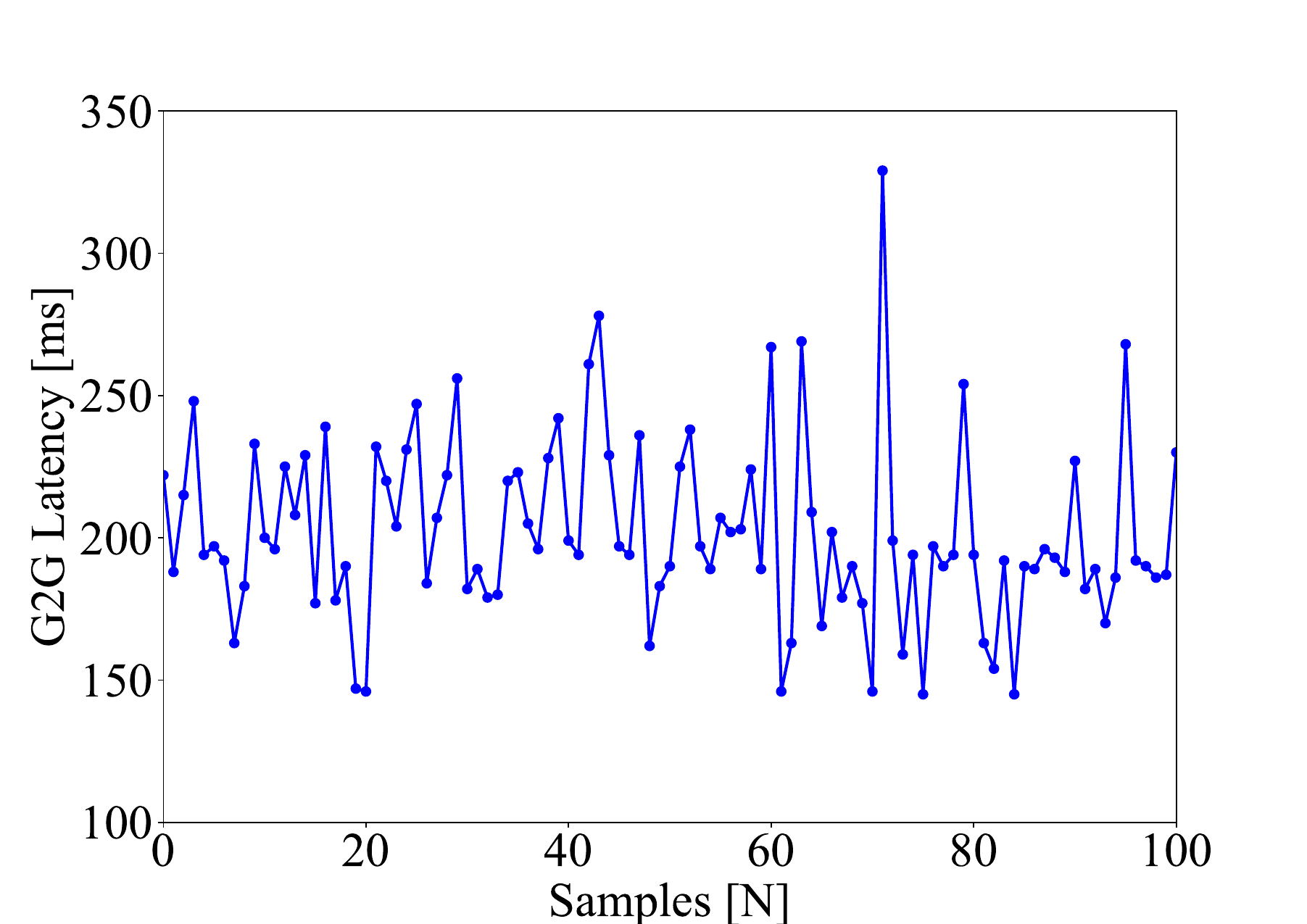}
    \caption{G2G latency between the vehicle camera and the teleoperated driving interface monitor. The achieved G2G latency enables safe teleoperated driving for low speed.}
    \label{fig:g2g}
\end{figure}

\begin{figure}[t]
    \centering
    \includegraphics[scale=0.255, trim={0cm 0cm 0cm 2.5cm},clip]{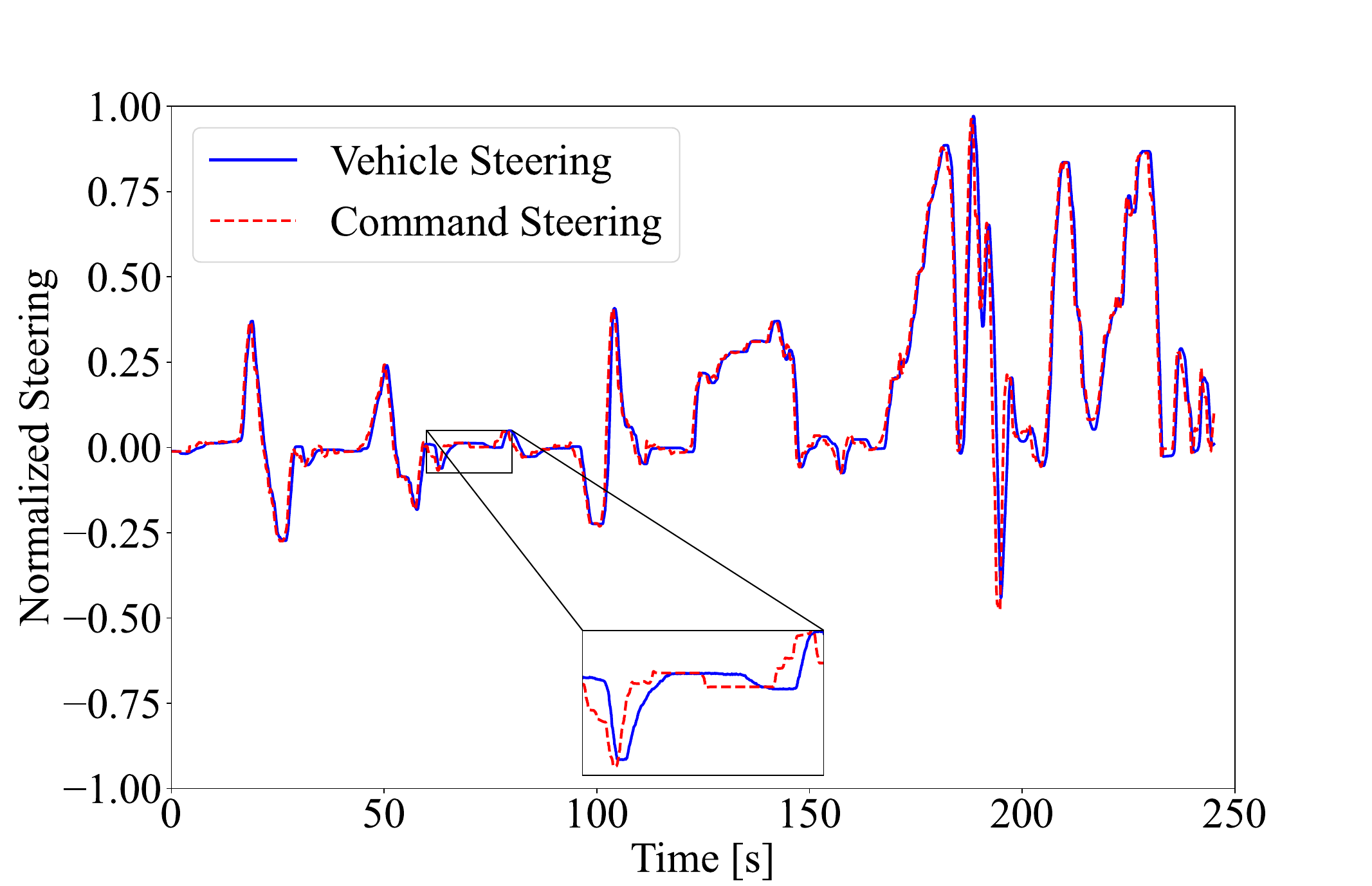}
    \caption{Realized vehicle steering angle and steering command sent from teleoperation interface, both normalized to [-1, 1].}
    \label{fig:steering_lat}
\end{figure}

\subsection{Network}
\label{sec:network-perf}
The network setup used for the experiments is depicted in Fig. \ref{fig:tod_arch}. The remote vehicle is connected to the internet via a 5G phone used as a modem, while the teleoperated driving interface is directly connected to the internet via an Ethernet interface. Both the teleoperated interface and the remote vehicle are connected to a VPN to be easily and securely reachable at the cost of network performance. All experiments were conducted in the Kirchberg district of Luxembourg City using a commercial non-standalone 5G network operating on the 3.6 GHz band. To evaluate network performance, we used iPerf to measure latency, jitter, and packet loss. We observed an RTT of 27.63 ms when transmitting TCP packets of 128 KB size, and a jitter of 0.709 ms when transmitting 85,068 UDP datagrams, with zero packet loss.
These results provide a quantitative assessment of network stability and its impact on real-time teleoperation performance.

\subsection{Glass to Glass Latency}
G2G latency is a crucial metric of teleoperation as it directly relates to the user experience of the system. A high G2G latency will strongly limit the teleoperation capabilites in terms of control, maximum speed and ultimately safety. Measuring G2G latency reliably and in a non-intrusive manner is not a trivial problem \cite{g2g}, in this work we decided to measure the G2G latency using a basic and straightforward method at the expense of some accuracy. A monitor showing a running clock is placed in front of the vehicle camera while another monitor connected to the teleoperation interface is placed next to the first monitor. The measurement is taken by recording a video of the two monitors and manually reporting the time difference between the clocks displayed on both monitors. The measurement camera and running clock monitor both run at their own 60Hz cadence, this alone could induce up to 17ms of error on individual measurements. Identification of digits in the sub-100ms range proved challenging, and therefore errors of up to 99ms could arise from those ambiguous readings. Overall, stemming from the mean error, we consider the accuracy of the measurements to be around 50ms. A total of 100 measurements over 1min and 30s of video were taken, the results are depicted in Fig. \ref{fig:g2g} and Table \ref{tab:results}. The average recorded G2G latency is 202.41ms with a standard deviation of 31.56ms. Fig. \ref{fig:g2g} shows that high G2G latencies can reach over 300ms while low values drop to 150ms. These results indicate that even on a 5G network, a live video stream can still induce a significant latency considering that, at 30km/h, a vehicle will cover approximately 1.7 meters. Integrating the possible measurement errors, we conclude that the achieved G2G latency can enable safe teleoperated driving operation at low speed (less than 20km/h). As the presented teleoperated driving system uses a relatively standard video streaming pipeline, the G2G latency could be further reduced by refining the streaming parameters or introducing novel methods \cite{7976319}.

\begin{table}[t!]
    \centering
    \caption{G2G latency and RTT measurements.}
    \begin{tabular}{c|c|c|c}
        \hline
        \hline
        & Number of samples & Average [ms] & Standard Deviation [ms]\\
        \hline
        G2G & 100 & 202.41 & 31.56\\
        \hline
        RTT & 1000 & 46.63 & 10.05\\
        \hline
        \hline
    \end{tabular}
    \label{tab:results}
\end{table}

\subsection{Round-Trip Time (RTT)}
Both the teleoperation commands and vehicle telemetry should be exchanged as fast as possible in order to provide the smoothest experience. The following experiment measures the RTT of a command sent from the teleoperated driving interface to the remote vehicle. A teleoperated driving command contains a timestamp that once received by vehicle will be sent back in a telemetry message. The RTT is measured as the time difference between the reception of a telemetry message and the corresponding teleoperation command timestamp included. The measurement is run over 1000 samples and the results are reported in Fig. \ref{fig:rtt} and Table \ref{tab:results}. The average measured RTT stands at 46.67ms with a standard deviation of 10.05ms. Fig. \ref{fig:rtt} shows punctual highs at 80 or even 120ms while punctual lows around 30ms can also be observed. These results are in line with the network performance measured in Section \ref{sec:network-perf} and indicate that the presented teleoperated driving solution does not induce significant network overhead. Furthermore, the measured RTT is also influenced by the underlying implementation and inner workings of the teleoperation system. The RoboCar teleoperated driving component, as previously mentioned, runs at 100hz and this can already contribute to up to 10ms of delay. Assuming that the average latency is half that of the RTT, a vehicle driving at 30km/h will cover a distance of around 0.2m over 23ms and therefore, while it would always be beneficial to reduce the RTT further, the achieved performance should allow for a safe teleoperated driving at moderate to high speeds.

\subsection{Steering Control}

Another important aspect of the user experience is the steering response behavior, which reflects how well the remote vehicle follows the commanded steering inputs. Unlike G2G latency and RTT, which are primarily influenced by network conditions, steering response is largely determined by the vehicle control system and steering dynamics. To qualitatively assess this behavior, we analyze the recorded steering input and actual vehicle response over data 5000 samples, corresponding to approximately 250 seconds of driving, as shown in Fig.~\ref{fig:steering_lat}.
While this analysis provides insights into the smoothness, stability, and overall precision of the teleoperated steering manoeuvres, it does help identifying any potential inconsistencies or delays introduced by the vehicle actuation system. Overall, the latency due to network and steering system time constant is low but can reach up to 750ms. Such a relatively big delay in realizing steering commands creates a cognitive load on the driver besides the other factors induced by the latency in video steaming. Future works may investigate and study in details these factors while simultaneously improving latencies and steering system performance. In light of this analysis, we consider the system safe to operate at low speed (less than 20km/h).

\section{Conclusion and Future Work}
\label{sec:conclusion}
This paper presented an experimental evaluation of teleoperated driving over 5G networks, focusing on network performance. Using a real-world testbed based on a Kia Soul EV equipped with a drive-by-wire system, we analyzed key teleoperated driving performance using standard evaluation metrics such as G2G latency and RTT in addition to a steering control delay evaluation. The experimental results of our teleoperated driving setup highlight the feasibility of 5G-enabled teleoperated driving while also identifying critical challenges to reach realistic teleoperated driving deployments.
While our study primarily focused on network performance, future work will explore additional factors that influence teleoperated driving reliability and usability. Specifically, we aim to investigate adaptive network optimization and streaming strategies to mitigate the effects of varying connectivity conditions. Furthermore, incorporating haptic feedback and predictive control mechanisms could enhance the teleoperation experience by compensating for minor network disruptions. Lastly, extending the evaluation to longer driving scenarios in diverse urban and rural environments will provide deeper insights into the real-world scalability of teleoperated driving over 5G. By addressing these challenges, future research can help refine teleoperated systems, bringing them closer to large-scale deployment in autonomous mobility solutions.

\section*{Acknowledgments}
This research was funded in whole, or in part, by the
Luxembourg National Research Fund (FNR). The authors would further like to thank all partners supported this initiative over the years.

\bibliographystyle{ieeetr}
\bibliography{bibliography}

\begin{thebibliography}{10}

\bibitem{zhao2024remote}
L.~Zhao, M.~Nybacka, M.~Aramrattana, M.~Rothh{\"a}mel, A.~Habibovic, L.~Drugge, and F.~Jiang, ``Remote driving of road vehicles: A survey of driving feedback, latency, support control, and real applications,'' {\em IEEE Transactions on Intelligent Vehicles}, 2024.

\bibitem{georg2018teleoperated}
J.-M. Georg, J.~Feiler, F.~Diermeyer, and M.~Lienkamp, ``Teleoperated driving, a key technology for automated driving? comparison of actual test drives with a head mounted display and conventional monitors,'' in {\em 2018 21st International Conference on Intelligent Transportation Systems (ITSC)}, pp.~3403--3408, IEEE, 2018.

\bibitem{tang2014teleoperated}
T.~Tang, F.~Chucholowski, and M.~Lienkamp, ``Teleoperated driving basics and system design,'' {\em ATZ worldwide}, vol.~116, no.~2, pp.~16--19, 2014.

\bibitem{10178441}
S.~Rafiei, C.~Singhal, K.~Brunnström, and M.~Sjöström, ``Human interaction in industrial tele-operated driving: Laboratory investigation,'' in {\em 2023 15th International Conference on Quality of Multimedia Experience (QoMEX)}, pp.~91--94, 2023.

\bibitem{g2g}
C.~Bachhuber and E.~Steinbach, ``A system for high precision glass-to-glass delay measurements in video communication,'' pp.~2132--2136, 09 2016.

\bibitem{hawlader2024leveraging}
F.~Hawlader, F.~Robinet, and R.~Frank, ``Leveraging the edge and cloud for v2x-based real-time object detection in autonomous driving,'' {\em Computer Communications}, vol.~213, pp.~372--381, 2024.

\bibitem{hawlader2025cloud}
F.~Hawlader, F.~Robinet, G.~Elghazaly, and R.~Frank, ``Cloud-assisted 360-degree 3d perception for autonomous vehicles using v2x communication and hybrid computing,'' in {\em 2025 20th Wireless On-Demand Network Systems and Services Conference (WONS)}, 2025.

\bibitem{testouri2025robocar}
M.~Testouri, G.~Elghazaly, and R.~Frank, ``Robocar: A rapidly deployable open source platform for autonomous driving research,'' {\em IEEE Intelligent Transportation Systems Magazine}, pp.~2--15, 2025.

\bibitem{cislaghi2023simulation}
V.~Cislaghi, C.~Quadri, V.~Mancuso, and M.~A. Marsan, ``Simulation of tele-operated driving over 5g using carla and omnet++,'' in {\em 2023 IEEE Vehicular Networking Conference (VNC)}, pp.~81--88, IEEE, 2023.

\bibitem{saez2023design}
J.~Saez-Perez, Q.~Wang, J.~M. Alcaraz-Calero, and J.~Garcia-Rodriguez, ``Design, implementation, and empirical validation of a framework for remote car driving using a commercial mobile network,'' {\em Sensors}, vol.~23, no.~3, p.~1671, 2023.

\bibitem{kloppel2023identifying}
M.~Kl{\"o}ppel-Gersdorf, A.~Bellanger, and T.~Otto, ``Identifying challenges in remote driving,'' in {\em International Conference on Smart Cities and Green ICT Systems}, pp.~146--166, Springer, 2023.

\bibitem{neumeier2019towards}
S.~Neumeier and C.~Facchi, ``Towards a driver support system for teleoperated driving,'' in {\em 2019 IEEE Intelligent Transportation Systems Conference (ITSC)}, pp.~4190--4196, IEEE, 2019.

\bibitem{kim2025development}
K.~Kim, Y.~Jeon, M.~Kim, S.~Lee, and S.-C. Kee, ``Development of tele-operated driving assistance system for autonomous vehicles with an open-source platform,'' {\em International Journal of Automotive Technology}, pp.~1--12, 2025.

\bibitem{schimpe2022open}
A.~Schimpe, J.~Feiler, S.~Hoffmann, D.~Majstorovi{\'c}, and F.~Diermeyer, ``Open source software for teleoperated driving,'' in {\em 2022 International Conference on Connected Vehicle and Expo (ICCVE)}, pp.~1--6, IEEE, 2022.

\bibitem{neumeier2019measuring}
S.~Neumeier, E.~A. Walelgne, V.~Bajpai, J.~Ott, and C.~Facchi, ``Measuring the feasibility of teleoperated driving in mobile networks,'' in {\em 2019 Network Traffic Measurement and Analysis Conference (TMA)}, pp.~113--120, IEEE, 2019.

\bibitem{shahbazi2018systematic}
M.~Shahbazi, S.~F. Atashzar, and R.~V. Patel, ``A systematic review of multilateral teleoperation systems,'' {\em IEEE transactions on haptics}, vol.~11, no.~3, pp.~338--356, 2018.

\bibitem{passenberg2010survey}
C.~Passenberg, A.~Peer, and M.~Buss, ``A survey of environment-, operator-, and task-adapted controllers for teleoperation systems,'' {\em Mechatronics}, vol.~20, no.~7, pp.~787--801, 2010.

\bibitem{arcara2002control}
P.~Arcara and C.~Melchiorri, ``Control schemes for teleoperation with time delay: A comparative study,'' {\em Robotics and Autonomous systems}, vol.~38, no.~1, pp.~49--64, 2002.

\bibitem{eusebi1998force}
L.~Eusebi and C.~Melchiorri, ``Force reflecting telemanipulators with time-delay: Stability analysis and control design,'' {\em IEEE Transactions on Robotics and Automation}, vol.~14, no.~4, pp.~635--640, 1998.

\bibitem{de2025bridging}
A.~De~La Rosa-Garcia, A.~Soto-Marrufo, D.~Luviano-Cruz, A.~Rodriguez-Ramirez, and F.~Garcia-Luna, ``Bridging remote operations and augmented reality: A survey of current trends,'' {\em IEEE Access}, 2025.

\bibitem{7976319}
C.~Bachhuber, E.~Steinbach, M.~Freundl, and M.~Reisslein, ``On the minimization of glass-to-glass and glass-to-algorithm delay in video communication,'' {\em IEEE Transactions on Multimedia}, vol.~20, no.~1, pp.~238--252, 2018.

\bibitem{prasad20235g}
B.~K. Prasad, S.~Adarsh, and M.~Shinde, ``5g based remote control of autonomous vehicle,'' in {\em 2023 IEEE 3rd Mysore Sub Section International Conference (MysuruCon)}, pp.~1--6, IEEE, 2023.

\bibitem{mahmood2023remote}
A.~Mahmood, S.~F. Abedin, M.~O’Nils, M.~Bergman, and M.~Gidlund, ``Remote-timber: an outlook for teleoperated forestry with first 5g measurements,'' {\em IEEE Industrial Electronics Magazine}, vol.~17, no.~3, pp.~42--53, 2023.

\bibitem{aliaga2004experimental}
I.~Aliaga, A.~Rubio, and E.~Sanchez, ``Experimental quantitative comparison of different control architectures for master-slave teleoperation,'' {\em IEEE transactions on control systems technology}, vol.~12, no.~1, pp.~2--11, 2004.

\bibitem{neumeier2018way}
S.~Neumeier, N.~Gay, C.~Dannheim, and C.~Facchi, ``On the way to autonomous vehicles teleoperated driving,'' in {\em AmE 2018-Automotive meets Electronics; 9th GMM-Symposium}, pp.~1--6, VDE, 2018.

\bibitem{gnatzig2013system}
S.~Gnatzig, F.~Chucholowski, T.~Tang, and M.~Lienkamp, ``A system design for teleoperated road vehicles,'' in {\em International Conference on Informatics in Control, Automation and Robotics}, vol.~2, pp.~231--238, SciTePress, 2013.

\bibitem{neumeier2019yet}
S.~Neumeier, M.~H{\"o}pp, and C.~Facchi, ``Yet another driving simulator openrouts3d: The driving simulator for teleoperated driving,'' in {\em 2019 IEEE International Conference on Connected Vehicles and Expo (ICCVE)}, pp.~1--6, IEEE, 2019.

\bibitem{kim2024teleoperated}
K.~Kim and S.-C. Kee, ``Teleoperated driving with virtual twin technology: A simulator-based approach,'' {\em World Electric Vehicle Journal}, vol.~15, no.~7, p.~311, 2024.

\bibitem{kashwani2024evaluation}
F.~Kashwani, B.~Hassan, P.-Y. Kong, M.~Khonji, and J.~Dias, ``Evaluation of predictive display for teleoperated driving using carla simulator,'' in {\em 2024 IEEE/RSJ International Conference on Intelligent Robots and Systems (IROS)}, pp.~12190--12195, IEEE, 2024.

\bibitem{varisteas2019junior}
G.~Varisteas and R.~Frank, ``Junior, a research platform for connected and automated driving,'' in {\em Proceedings of 2019 IEEE Vehicular Networking Conference}, 2019.

\bibitem{oscc}
``Open source car control.'' \url{https://https://github.com/PolySync/OSCC}.
\newblock Accessed: 09-01-2024.

\bibitem{Hintjens2013ZeroMQMF}
P.~Hintjens, ``Zeromq: Messaging for many applications,'' 2013.

\bibitem{schulzrinne1998real}
H.~Schulzrinne, A.~Rao, and R.~Lanphier, ``Real time streaming protocol (rtsp),'' tech. rep., 1998.

\end{thebibliography}

\end{document}